\documentclass[a4paper]{svmult}
\usepackage{times}
\usepackage[dvips]{graphicx,epsfig}
\usepackage{amsmath,amsfonts,amssymb}
\usepackage{cite,url}
\newcommand{\bmit}[1]{\mbox{\boldmath $#1$}}

\begin{document}

\title*{Degeneracies in a nonintegrable pairing model}

\author{{J.~Oko{\l}owicz}\inst{1}
\and {{M.~P{\l}oszajczak}}\inst{2}
\and {J.~Dukelsky}\inst{3}}

\institute{{ Institute of Nuclear Physics, Radzikowskiego 152, PL-31342 Krak\'ow, Poland}
\and {Grand Acc\'{e}l\'{e}rateur National d'Ions Lourds (GANIL), CEA/DSM -- CNRS/IN2P3, \\
BP 5027, F-14076 Caen Cedex 05, France}
\and {Instituto de Estructura de la Materia, CSIC, Serrano 123, 28006 Madrid, Spain}
}

\maketitle

\begin{abstract}
The evolution pattern of exceptional points is studied in a non-integrable limit of the 
complex-extended 3-level Richardson-Gaudin model. The appearance of a pseudo-diabolic point from the fusion of two exceptional points is demonstrated in the anti-hermitian limit of the model and studied in some details.
\end{abstract}

\section{Introduction}

Small quantum systems, whose properties are profoundly affected by
environment, i.e., continuum of scattering and decay channels, are
intensely studied in various fields of physics (nuclear physics, atomic
and molecular physics, nanoscience, quantum optics, etc.). These
different open quantum systems (OQS), in spite of their specific
features, have generic properties, which are common to all weakly
bound/unbound systems close to the threshold. 
An essential part of the motion of short-lived nucleonic matter is in
classically forbidden regions, and their properties are impacted by both the continuum and many-body correlations \cite{Dob07,[Oko03]}.

Resonances are commonly found in various quantum systems, independently
of  their building blocks. Resonances are genuine intrinsic properties of
quantum systems describing preferential decays of  unbound states. The
effect of resonances and the non-resonant scattering states can be
considered in the OQS extension of the shell model (SM), the so-called
continuum shell model  (CSM) \cite{[Oko03]}.
Two realizations of the CSM have been studied recently: the real-energy CSM 
\cite{SMEC,[Oko03],SMEC_2p} and the  complex-energy CSM \cite{Bet02,Mic02} based on the Berggren ensemble \cite{Ber68}, the so-called Gamow Shell Model (GSM). 

For hermitian Hamilton operators, both real-energy CSM  \cite{SMEC,[Oko03],SMEC_2p} and complex-energy CSM (GSM)  \cite{Bet02,Mic02} lead naturally to the non-hermitian (complex-symmetric) eigenvalue problem. As a result, OQSs exhibit several unintuitive properties, which make them qualitatively different from  closed quantum systems (CQS). Among those salient features are: the segregation of time scales in the continuum \cite{Kle85,Dro00} (see also Ref. \cite{[Oko03]} for a recent review), the alignment of near-threshold states with decay channels \cite{Cha06,Dob07,[Oko03]}, the instability of SM eigenstates at the channel threshold \cite{Mic07,Oko08}, or the resonance crossings \cite{Rot00,[Oko03]}. In this lecture, we shall concentrate on the latter phenomenon which will be illustrated on the example of a non-integrable pairing model.

Most studies of degeneracies associated with avoided crossings in quantal spectra, focused on  the topological structure of the Hilbert space and the geometric phases \cite{Ber84,Lau94}. Among these degeneracies, one finds a diabolic point (DP) in hermitian Hamiltonians \cite{Ber84,Ber84a}, and an
exceptional point (EP) \cite{Kat95,Zir83,Hei91}, which appears in the complex $g$-plane 
of a generic Hamiltonian $H(g)=H_0+g H_1$, where both $H_0$ and $H_1$ are hermitian 
and $[H_0,H_1] \neq 0$. Below, we shall introduce a prototypical OQS, the 3-level Richardson-Gaudin (RG) model, to discuss the appearance of resonance crossings and their evolution with a parameter of a non-integrable perturbation. In particular, we shall show the appearance of a new kind of degeneracy, a pseudo-diabolic point (pseudo-DP) in the {\em anti-hermitian} limit of this model.

\section[]{The 3-level pairing model}

RG models \cite{Duk04} are based on the $SU(2)$ algebra with elements 
$K^+_l$, $K^-_l$, and $K^0_l$, 
fulfilling the commutation relations: $[K^+_l,K^-_{l^{\prime}}]=\delta_{ll^{\prime}} K^0_l~$ ,
$[K^0_l,K^{\pm}_{l^{\prime}}]=\pm\delta_{ll^{\prime}} K^{\pm}_l~$, where indices 
$l,l^{\prime}$ refer to a particular copy from a set of $L$, $SU(2)$ algebras.
 Each $SU(2)$ algebra possesses one quantum degree of freedom. In the following, we shall use the pair representation of SU(2) algebra leading to pairing 
Hamiltonians. The elementary operators in this representation are the number operators 
$N_j$ and the pair operators $A^{\dagger}_j$, $A_j$, defined as:
\begin{eqnarray}
N_{j}=\sum_{m}a_{jm}^{\dagger}a_{jm} ~;~~~~~
A_{j}^{\dagger}=\sum_{m}a_{jm}^{\dagger}a_{j\overline{m}}^{\dagger}=(A_{j}^{\dagger})^{\dagger}%
\label{eq2}
\end{eqnarray}
where $j$ is the total angular momentum and $m$ is the $z$-projection. The state 
${j\overline{m}}$ is the time reversal of ${jm}$. The relation between the operators of the 
pair algebra and the generators of the SU(2) algebra is:
\begin{eqnarray}
K_{l}^{0}=\frac{1}{2}N_{l}-\frac{1}{4}\Omega _{l}~;~~~~  K_{l}^{+}& =\left( K_{l}^{-}\right) ^{\dagger
}=\frac{1}{2}A_{l}^{\dagger } \label{K0}
\end{eqnarray}
where $\Omega_l$  is the particle degeneracy of level $l$. With this correspondence, one 
can introduce an integrable 3-level pairing Hamiltonian as:
\begin{eqnarray}
H_I(g)=\sum_{i}\varepsilon_{i}N_{i}+g\sum_{ij}A_{i}^{\dagger}A_j
\label{IPH}
\end{eqnarray}
Below, we shall consider the non-integrable version of the 3-level pairing model \cite{Duk09}:
\begin{eqnarray}
H_{NI}(g)=H_I(g)+g'\sum_{i}N_i^2~~~~, ~~~~ g'=\gamma g
\label{NIPH}
\end{eqnarray}
where $\gamma<0$ is a $C$-number.

\section[]{Level degeneracies}

The position of all possible degeneracies in the complex g-plane are indicated by the roots of the coupled equations:
\begin{eqnarray}
{\rm det}\left[H\left(  g\right)  -EI\right]  = 0~;~~~~ \frac{\partial}{\partial E} {\rm det}\left[ H\left(
g\right)  -EI\right] = 0
\end{eqnarray}
By eliminating $E$ from these two equations, we are left with the discriminant $D(g)$, 
a polynomial in $g$ of degree  $M=n(n-1)$, where $n$ is the number of eigenstates. The discriminant can be written as \cite{Zir83}:
\begin{eqnarray}
D(g)=\prod_{m<m^{\prime}}\left[E_m(g)-E_{m^{\prime}}(g)\right]^2
\end{eqnarray}
where $E_m(g)$, $E_{m^{\prime}}(g)$ denote the complex eigenvalues of $H(g)$. 
The eigenvalue degeneracies $E_m(g)=E_{m^{\prime}}(g)$ at $g=g_{\alpha}$ ($\alpha=1,\dots,M$), 
can be found numerically by looking for sharp minima of the functional $D(g)$. For the non-integrable Hamiltonian, the degenerate eigenvalues are either the single-root (EP) or  double-root solutions such as DP or pseudo-DP. 

\begin{figure}[t]\centering
\includegraphics[width=75mm,angle=-90]{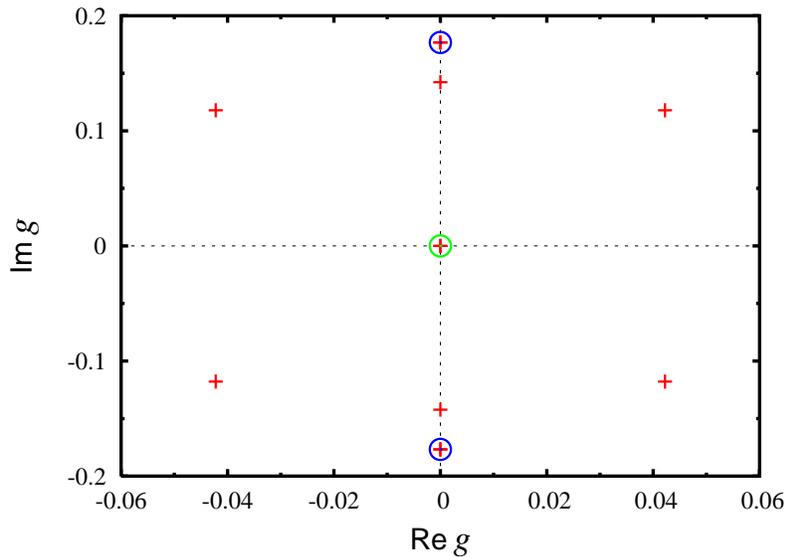}
\caption{Localization of EPs (the red cross) and pseudo-DPs (the red cross inside a blue circle) in the complex $g$-plane for the non-integrable 3-level pairing Hamiltonian with 
$g'/g=-1/2$ (see Eq. (\ref{NIPH})). For chosen energies of levels, one finds a trivial degeneracy at $g=0$ (the non-interacting limit). For more details, see the description in the text.}
\label{fig22}
\end{figure}
%
\begin{figure}[t]\centering
\includegraphics[width=75mm,angle=-90]{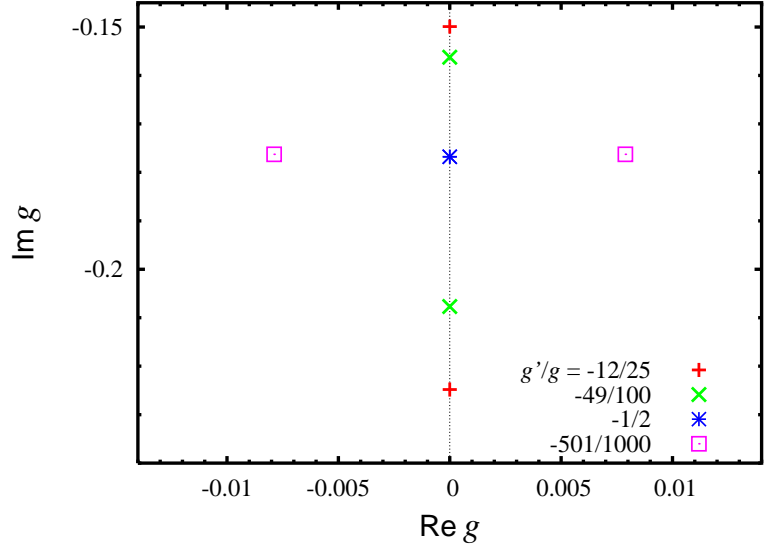}
\caption{The formation and decay of a pseudo-DP at $g'=-g/2$ (the blue star) in the anti-hermitian limit of the non-integrable 3-level pairing Hamiltonian (\ref{NIPH}).}
\label{fig23}
\end{figure}
%
\begin{figure}[t]\centering
\includegraphics[width=65mm,angle=-90]{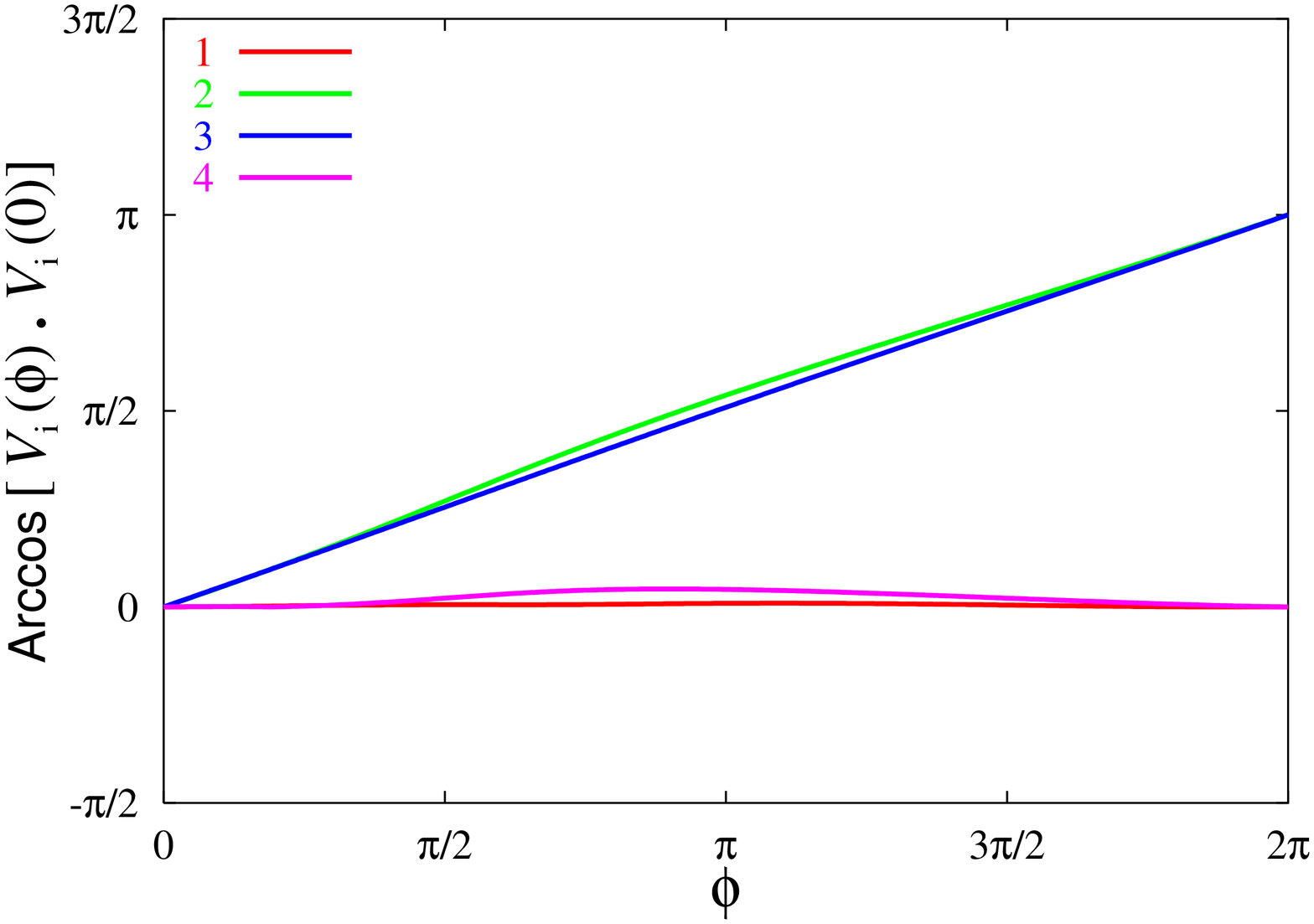}
\includegraphics[width=65mm,angle=-90]{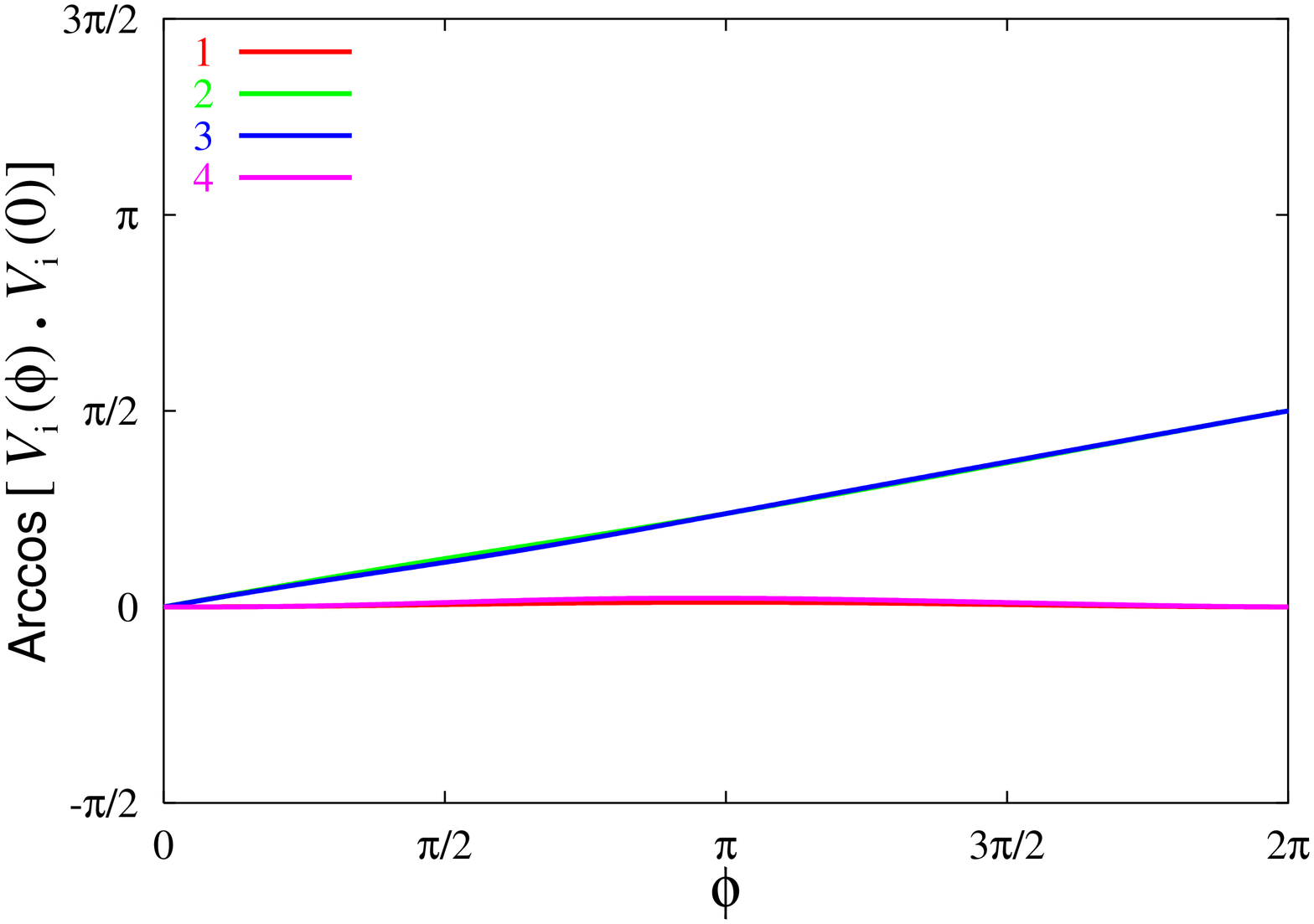}
\caption{The real parts of phase trajectories for four eigenvectors involved in the pseudo-DP at $g'/g=-1/2$ (the upper part). This degeneracy of two eigenvalues results from a coalescence of two EPs. In the lower part, the real parts of phase trajectories for two eigenvectors which form an EP at $g'/g=-49/100$ (see Fig. \ref{fig23}) are shown. The phase of each vector is defined  with respect of its reference value at $\varphi =0$.} 
\label{fig24}
\end{figure}
%
\begin{figure}[t]\centering
\includegraphics[width=65mm,angle=-90]{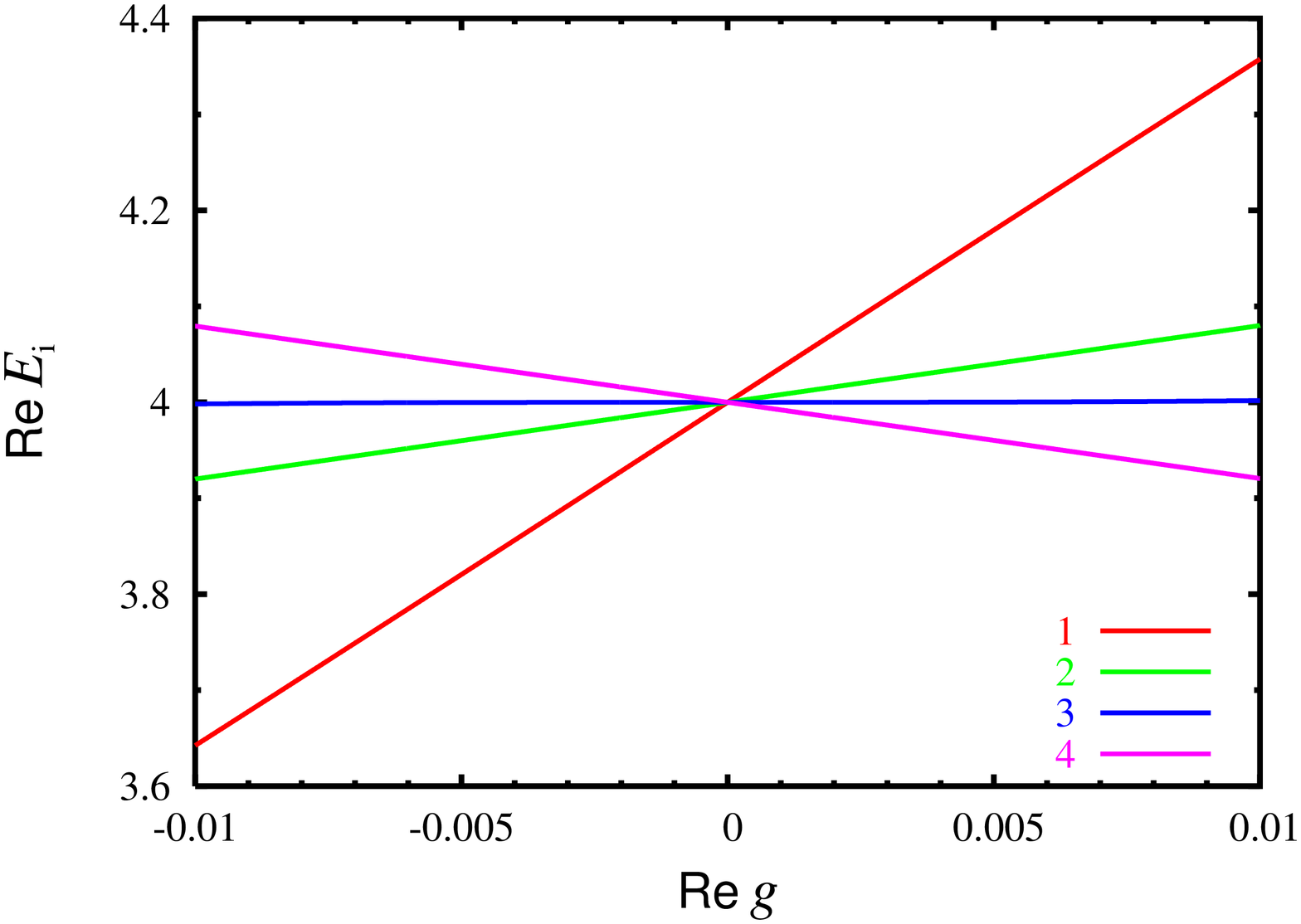}
\includegraphics[width=65mm,angle=-90]{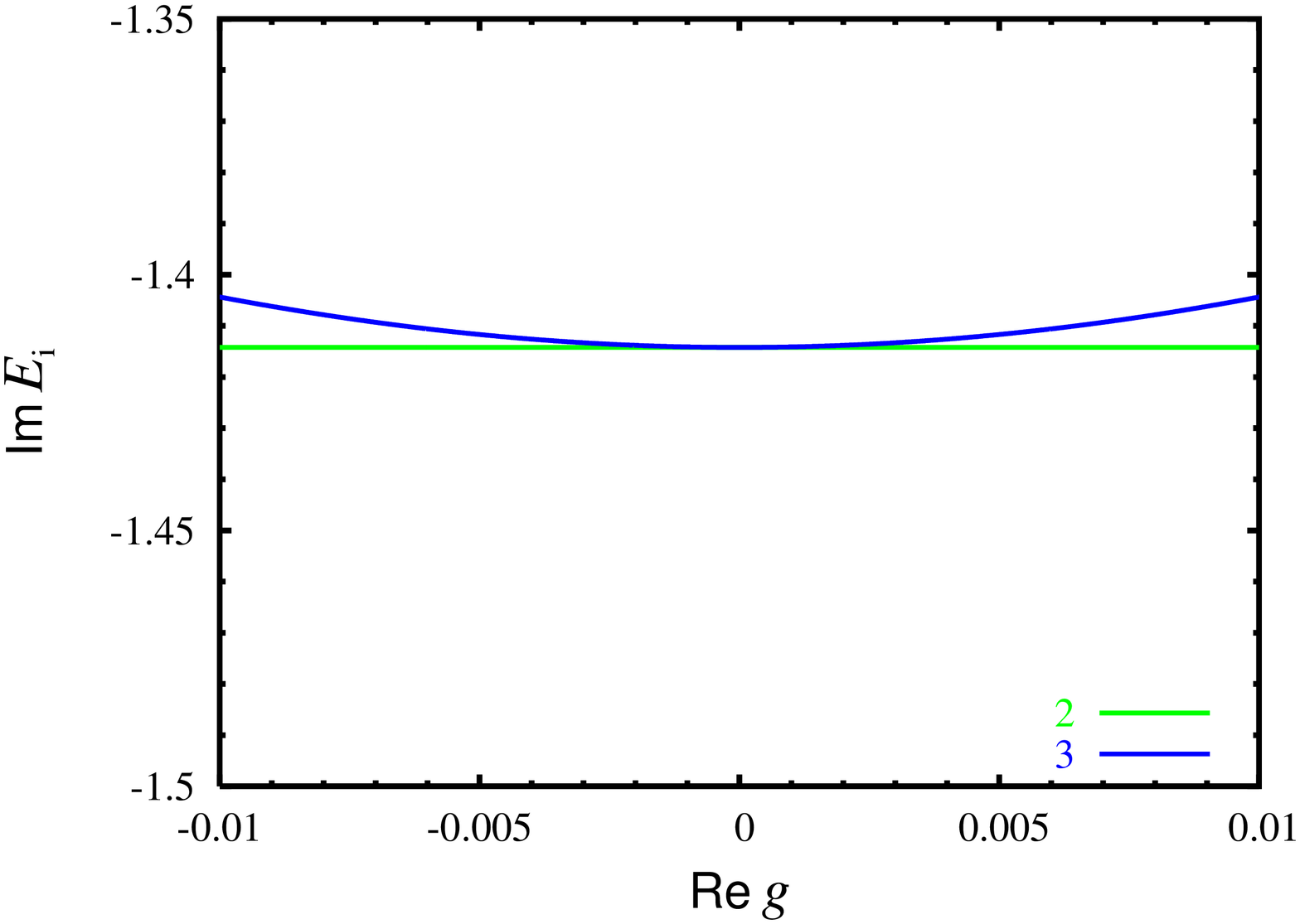}
\caption{Real and imaginary parts of eigenvalues at around the pseudo-DP 
$g=-{\rm i}/(4\sqrt{2})$ are plotted along the cut 
$({\cal R}e(g),-1/(4\sqrt{2}))$ in $g$-plane. For more details, see the description in the text.}
 \label{fig25}
\end{figure}
%
\begin{figure}[t]\centering
\includegraphics[width=75mm,angle=-90]{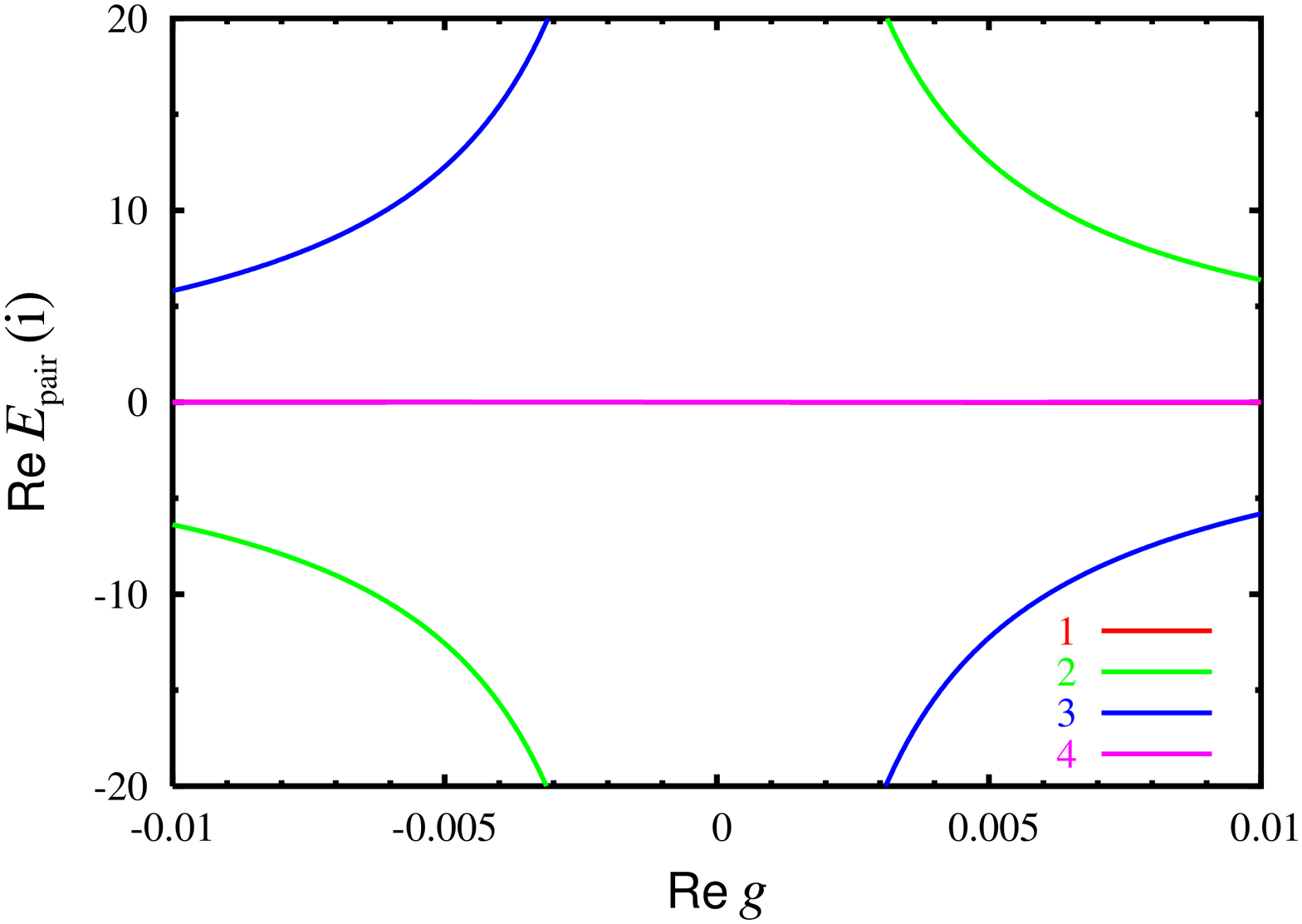}
\caption{The pairing energy at around the pseudo-DP of the non-integrable 3-level pairing Hamiltonian ($g'=-g/2$). }
\end{figure}

\section[]{Properties of a pseudo-diabolic point in the anti-hermitian limit of a non-integrable 3-level pairing model}

Let us solve the non-integrable 3-level pairing model for the case of 2 pairs of fermions in a valence space of degeneracy $\Omega_1=2, \Omega_2=6, \Omega_3=2$. The energies of three levels are 
$\varepsilon _{1}=0$, $\varepsilon _{2}=1$, and $\varepsilon _{3}=2$. In this model space, there are 4 many-body states.

Fig. \ref{fig22} shows the global pattern of level crossings in the complex $g$-plane for 
$g'/g=-1/2$.  In the lower half-plane of $g$, all eigenvalues are either discrete states at the real $g$-axis or decaying resonances. Complex conjugate degeneracies situated in the upper half-plane 
(${\cal I}m(g)>0$) correspond to capturing resonances. 

Fig. \ref{fig23} shows a typical pattern associated with the formation/breakup of the pseudo-DP at 
$g=-{\rm i}/(4\sqrt{2})$. With an increasing value of $g'/g$, the two EPs in the complex $g$-plane approach each other, coalesce at $g'/g=-1/2$ and, subsequently, move along the 
axis ${\cal R}e(g)=0$ for $g'/g>-1/2$. The coalescence of two EPs at $g'/g=-1/2$, leads to the 
formation of a double-root singularity,  for which the geometric phase associated with a cyclic 
evolution changes twice faster than for the EP. On the other hand and in contrast to the ordinary DP, both eigenvalues {\em and} eigenvectors merge, like for an ordinary EP. 

The upper part of Fig. \ref{fig24} shows real parts of phase trajectories around the pseudo-DP for all four eigenvectors in the model space.  At the pseudo-DP, real part of their eigenenergies are identical. 
These trajectories in $g$-plane are described by: 
$g = -(1/(4\sqrt{2}){\rm i} + 0.01{\rm e}^{{\rm i}\varphi}$, 
where $\varphi $ defines the way of encircling the degeneracy. The phase of each vector is defined 
with respect of its value at $\varphi =0$. The eigenvalues for vectors 
'2' and '3' are degenerate at the pseudo-DP and change their phases 
by $\pi $ at each encircling. Simultaneously, phases of  vectors for non-degenerate eigenvalues '1' and '4' remain approximately constant. For a comparison, the lower part of Fig. \ref{fig24} exhibits real parts of phase trajectories of all four vectors around the EP $g=-0.207687{\rm i}$ at $g'/g=-49/100$ (see Fig. \ref{fig23}).  These trajectories are specified by: $g = -0.207687{\rm i} + 0.01{\rm e}^{{\rm i}\varphi}$. 
Eigenvalues for vectors '2' and '3' are degenerate at the EP. After first encircling of the EP they become exchanged. After a second encircling, the phases of vectors  '2' and '3' change by $\pi $, and after two more encircling of the EP all phases return to their initial values.

The behavior of eigenenergies in the neighborhood of a pseudo-DP in complex $g$-plane are shown along the cut $({\cal R}e(g),-1/(4\sqrt{2}))$ (see Fig. \ref{fig25}). The fusion of two EPs and the formation of a single pseudo-DP is seen as a sharp crossing of four energies and a coalescence of two widths (the eigenvectors '2' and '3'). Even though phases of eigenvectors '1' and '4' are almost constant when encircling the pseudo-DP (see Fig. \ref{fig24}), they are essential for the formation of a pseudo-DP degeneracy. It can be shown rigorously, that the pseudo-DP does not result from the Jordan block of 2-level system.

\section[]{Mixing of wave functions at the pseudo-diabolic point: the pairing correlation energy}

Salient features of eigenvectors around a pseudo-DP of a 3-level pairing Hamiltonian (\ref{NIPH})  with  
$g'/g=-1/2$ can be studied analytically. Let $g=- {\rm i}(4\sqrt{2})^{-1}+\delta$, where $\delta$ is a complex number. The eigenenergies of a Hamiltonian matrix:
\begin{eqnarray}
\left(\begin{matrix} E_1 \cr E_2 \cr E_3 \cr E_4 \cr \end{matrix}\right) =
\left(\begin{matrix}  
     4 - 3.79878\,\imag  + 35.9338\,\delta + {{\cal O}(\delta)}^2 \cr 
     4 - \sqrt{2}\,\imag  + 8\,\delta + {{\cal O}(\delta)}^2 \cr 
     4 - \sqrt{2}\,\imag  + {{\cal O}(\delta)}^2 \cr 
     4 + 0.263243\,\imag  - 7.93378\,\delta + {{\cal O}(\delta)}^2 \cr 
\end{matrix}\right) 
\label{eigen}
\end{eqnarray}
depend parametrically on the parameter $\delta$. Notice an asymmetric dependence on $\delta$ of eigenvalues '2' and '3' which form a pseudo-DP.

The eigenvectors, which are normalized according to the dual metric of RHS \cite{Ber68,Gel61,Mau68,Boh78}: 
\begin{eqnarray}
< {\hat{\bmit u}}_i|{\tilde {\hat{\bmit u}}}_j >= \delta_{ij}
\end{eqnarray} 
can be written as:
\begin{eqnarray}
{\hat {\bmit u}}_1 &=& 
  \left[ \begin{matrix}0.616894 + 0.517406\,\imag  + 
    \left( 7.05076 - 1.17993\,\imag  \right) \,\delta - 
    \left( 27.8062 + 130.27\,\imag  \right) \,{\delta}^2 + 
    {{\cal O}(\delta)}^3 \cr 0.731723 - 5.83203\,\imag \,\delta - 
    127.824\,{\delta}^2 + {{\cal O}(\delta)}^3 \cr 
   0.488757 - 3.21838\,\imag \,\delta - 67.7276\,{\delta}^2 + 
    {{\cal O}(\delta)}^3 \cr 0.616894 - 0.517406\,\imag  - 
    \left( 7.05076 + 1.17993\,\imag  \right) \,\delta - 
    \left( 27.8062 - 130.27\,\imag  \right) \,{\delta}^2 + 
    {{\cal O}(\delta)}^3\end{matrix} \right] ,
\nonumber \\ 
{\hat {\bmit u}}_2 &=& \left[ \begin{matrix}
    -\frac{ (1 - \imag )}{4\,\sqrt{2\sqrt{2}}}\,\frac{1}{{\sqrt{\delta}}} - 
    3\,\frac{ 1 + \imag }{4\,\sqrt{\sqrt{2}}}\,{\sqrt{\delta}} + 
    {{\cal O}(\delta)}^{\frac{3}{2}} \cr 
   \frac{1 + \imag }{4\,\sqrt{\sqrt{2}}}\frac{1}{{\sqrt{\delta}}} + 
    \frac{1 - \imag }{2\,\sqrt{2\sqrt{2}}}\,{\sqrt{\delta}} + 
    {{\cal O}(\delta)}^{\frac{3}{2}} \cr 
   {{\cal O}(\delta)}^{\frac{3}{2}} \cr
   \frac{1 - \imag }{4\,\sqrt{2\sqrt{2}}}\,\frac{1}{{\sqrt{\delta}}} + 
    3\,\frac{1 + \imag }{4\,\sqrt{\sqrt{2}}}\,{\sqrt{\delta}} + 
    {{\cal O}(\delta)}^{\frac{3}{2}}\end{matrix} \right] ,
\nonumber \\ 
{\hat {\bmit u}}_3 &=& 
  \left[ \begin{matrix}
    -\frac{1 + \imag }{4\,\sqrt{2\,\sqrt{2}}}\,\frac{1}{{\sqrt{\delta}}} - 
    \frac{9\,\sqrt{2} - 8 - \left( 9\,\sqrt{2} + 8 \right) \,\imag}
         {8\,\sqrt{2\,\sqrt{2}}}\,{\sqrt{\delta}} + 
    {{\cal O}(\delta)}^{\frac{3}{2}} \cr 
   -\frac{1 - \imag }{4\,2^{1/4}}\,\frac{1}{{\sqrt{\delta}}} + 
    \frac{1 + \imag }{4\,\sqrt{2\,\sqrt{2}}} \,{\sqrt{\delta}} + 
    {{\cal O}(\delta)}^{\frac{3}{2}} \cr 
   \sqrt{3\,\sqrt{2}}\,\left( 1 + \imag  \right) \,{\sqrt{\delta}} + 
    11\,\sqrt{\frac{3\,\sqrt{2}}{2}}\,\left( 1 - \imag  \right) 
                    \,{\delta}^{\frac{3}{2}} + 
    {{\cal O}(\delta)}^{\frac{5}{2}} \cr 
   \frac{1 + \imag }{4\,\sqrt{2\,\sqrt{2}}}\,\frac{1}{{\sqrt{\delta}}} + 
    \frac{9\,\sqrt{2} + 8 - \left( 9\,\sqrt{2} - 8 \right) \,\imag}
         {8\,\sqrt{2\,\sqrt{2}}}\,{\sqrt{\delta}} + 
    {{\cal O}(\delta)}^{\frac{3}{2}}\end{matrix} \right] ,
\nonumber \\ 
{\hat {\bmit u}}_4 &=& 
  \left[ \begin{matrix}0.345603 - 0.412057\,\imag  - 
    \left( 2.62262 - 1.08314\,\imag  \right) \,\delta + 
    \left( 14.2829 + 10.4398\,\imag  \right) \,{\delta}^2 + 
    {{\cal O}(\delta)}^3 \cr -0.582736 + 0.412499\,\imag \,\delta - 
    6.21696\,{\delta}^2 + {{\cal O}(\delta)}^3 \cr 
   -0.87242 + 3.06004\,\imag \,\delta + 26.4056\,{\delta}^2 + 
    {{\cal O}(\delta)}^3 \cr 0.345603 + 0.412057\,\imag  + 
    \left( 2.62262 + 1.08314\,\imag  \right) \,\delta + 
    \left( 14.2829 - 10.4398\,\imag  \right) \,{\delta}^2 + 
    {{\cal O}(\delta)}^3\end{matrix} \right] 
\nonumber 
\end{eqnarray}

In the limit $\delta \rightarrow 0$, the eigenvectors ${\hat {\bmit u}}_2$ and ${\hat {\bmit u}_3}$ with (identical) eigenvalues $E_2, E_3$ (see Eq. (\ref{eigen})) have divergent components. This feature of eigenvectors at a pseudo-DP makes it similar to an ordinary EP and leads to a singular behavior of various quantities. 

Let us consider for example the pairing operator: $g\sum_{ij}A_{i}^{\dagger}A_j$, in the neighborhood of  a pseudo-DP for $g'/g=-1/2$: 
\begin{eqnarray}
\left(
\begin{matrix} a_2\,\imag  + b_2\,\delta + {{\cal O}(\delta)}^2 & 
a_5/\sqrt{\delta} + {\cal O}(\delta)^{1/2} & 
a_6/\sqrt{\delta} + {\cal O}(\delta)^{1/2} & 
a_4\,\imag  + b_4\,\delta + {{\cal O}(\delta)}^2 \cr 
a_5/\sqrt{\delta} + {\cal O}(\delta)^{1/2} & 
a_1/{\delta} + {{\cal O}(\delta)}^0 & 
a_1\,\imag /{\delta} + {{\cal O}(\delta)}^0 & 
a_7/\sqrt{\delta} + {\cal O}(\delta)^{1/2} \cr 
a_6/\sqrt{\delta} + {\cal O}(\delta)^{1/2} & 
a_1\,\imag /{\delta} + {{\cal O}(\delta)}^0 & 
-a_1/{\delta} + {{\cal O}(\delta)}^0 &
a_8/\sqrt{\delta} + {\cal O}(\delta)^{1/2} \cr 
a_4\,\imag  +b_4\,\delta + {{\cal O}(\delta)}^2 & 
a_7/\sqrt{\delta} + {\cal O}(\delta)^{1/2} & 
a_8/\sqrt{\delta} + {\cal O}(\delta)^{1/2}  & 
a_3\,\imag  + b_3\,\delta + {{\cal O}(\delta)}^2 \cr  \end{matrix} 
\right) 
\nonumber 
\end{eqnarray}
Coefficients $a_i$ in this expression are: 
\begin{eqnarray}
&& 
a_1 = \frac{1}{16}, \, 
a_2 =-7.43796, \, b_2 =-3.64018, 
\nonumber \\ && 
a_3 =0.455281, \, b_3 =0.765176, \, 
a_4 =0.603023, \, b_4 =1.8219, 
\nonumber \\ && 
a_5 =0.475579\,(1 - \imag ), \, b_5 =0.449184\,(1 + \imag ), \, 
a_6 =0.475579\,(1 + \imag ), \, b_6 =4.59509\,(1 - \imag ), 
\nonumber \\ && 
a_7 =0.118873\,(1 - \imag ), \, b_7 =1.20969\,(1 + \imag ), \, 
a_8 =0.118873\,(1 + \imag ), \, b_8 =0.0511439\,(1 - \imag ) 
\nonumber
\end{eqnarray}

It is readily seen that all matrix elements of the pairing operator involving states '2' and '3' exhibit the square-root divergence if $\delta \rightarrow 0$, i.e. when approaching a pseudo-DP. This divergence is cancelled out if one adds diagonal pairing matrix elements for these two states. However, off-diagonal matrix elements between 'regular' states  (states '1' and '4' in this case) and any combination of states '2' and '3' are divergent because: $a_5 = {a_6}^\star$ and $a_7 = {a_8}^\star$. 

Dependence of a real part of the pairing energy on the interaction strength $g$ in the non-integrable pairing Hamiltonian (\ref{NIPH}) with $g'/g=-1/2$ is plotted along the cut at $({\cal R}e(g),-1/(4\sqrt{2}))$. Pairing energies in strongly mixed eigenstates '2' and '3' have opposite signs and diverge when a pseudo-DP is approached, i.e. if ${\cal R}e(g)\rightarrow 0$. On the other hand, the sum of pairing energies in states '2' and '3'   is both positive and finite and changes smoothly close to a pseudo-DP. These features of pairing energy in the neighborhood of a pseudo-DP are essentially different from those discussed in connection with effects of an ordinary DP on pair-transfer amplitude \cite{DP} as the coupling between eigenvectors involved vanishes at the DP.

\section[]{Conclusions}

Besides accidental degeneracies, sharp level crossings in the hermitian eigenvalue problem are allowed only between states of different quantum numbers (different symmetries). For the non-hermitian (complex-symmetric) eigenvalue problem, such as found in OQSs, exact eigenvalue degeneracies appear also for states of the same quantum numbers/symmetries. These are: (i) the DP, when Riemann sheets on which live eigenvalues just touch each other \cite{Ber84}, and (ii) the EP, when two Riemann sheets are entangled by the square-root type of singularity \cite{Kat95,Zir83,Hei91}. We have analyzed here features of degeneracies in the non-hermitian and non-integrable 3-level pairing model (see Eq. (\ref{NIPH}) which is a schematic representation of the OQS Hamiltonian. In particular, we have shown the appearance of a new kind of degeneracy, the pseudo-DP, in  the anti-hermitian limit of this model as a result of the coalescence between two EPs. The resonance eigenfunctions at the pseudo-DP (EP) are entangled and, except for the (complex) eigenenergy, it is impossible to define any physical quantity separately in each of the two resonances involved in the pseudo-DP degeneracy. 

This entanglement of wave functions may involve many states of an OQS, leading to strong mixing of wave functions for states of largely different eigenenergies. This mechanism is efficient not only in the close neighborhood of a particular pseudo-DP (EP), i.e. for a particular  choice of coupling constants in the OQS Hamiltonian, but also for systems with largely different coupling constants. Future studies, using the CSM and realistic effective interactions could provide a first attempt to provide a consistent description of continuum wave functions and related observables in the presence of a pseudo-DP (EP).

\section*{Acknowledgments}

One of us (M.P.) wish to thank Witek Nazarewicz for useful discussions. This work was supported in part by the Spanish MEC under grant No. FIS2006-12783-C03-01 and by the CICYT(Spanish)-IN2P3(French) cooperation.


\begin{thebibliography}{99}
\bibitem{Dob07} J. Dobaczewski, N. Michel, W. Nazarewicz, M. P{\l}oszajczak, and
J. Rotureau, {\em Prog. Part. Nucl. Phys.} {\bf 59}, 432 (2007).
\bibitem{[Oko03]}
J. Oko{\l}owicz, M. P{\l}oszajczak, and I. Rotter, {\em Phys. Rep.} {\bf 374}, 271 (2003).
\bibitem{SMEC} K.~Bennaceur, F.~Nowacki, J.~Oko{\l}owicz, and M.~P{\l}oszajczak, {\em Nucl. Phys.} A {\bf 651}, 289 (1999) ; {\em Nucl. Phys.} A {\bf 671}, 203 (2000). 
\bibitem{SMEC_2p} J.~Rotureau, J.~Oko{\l}owicz, and M.~P{\l}oszajczak, {\em Phys. Rev. Lett.} {\bf 95}, 042503 (2005) ; {\em Nucl. Phys.} A {\bf 767}, 13 (2006).
\bibitem{Ber68} T.~Berggren, {\em Nucl. Phys.} A {\bf 109}, 265 (1968). 
\bibitem{Bet02} R. ~Id Betan, R.J. ~Liotta, N. ~Sandulescu, and T.~Vertse, {\em Phys. Rev. Lett.} {\bf 89},  042501 (2002). 
\bibitem{Mic02} N.~Michel, W.~Nazarewicz, M.~P{\l}oszajczak, and K. Bennaceur,
{\em Phys. Rev. Lett.} {\bf 89}, 042502 (2002);\\ 
N.~Michel, W.~Nazarewicz, M.~P{\l}oszajczak, and J.~Oko{\l}owicz, {\em Phys. Rev.} C {\bf 67}, 054311 (2003);\\ 
N.~Michel, W.~Nazarewicz, and M.~P{\l}oszajczak, {\em Phys. Rev.} C {\bf 70}, 064313 (2004). 
\bibitem{Gel61} I.M. Gel'fand and N.Ya. Vilenkin, 
{\em Generalized Functions}, Vol. {\bf 4}, Academic Press, New York (1961).
\bibitem{Mau68} K. Maurin, {\em Generalized Eigenfunction Expansions and Unitary Representations of Topological Groups}, Polish Scientific Publishers, Warsaw (1968).
 \bibitem{Boh78} A. Bohm, {\em The Rigged Hilbert Space and Quantum Mechanics},
 Lecture Notes in Physics {\bf 78}, Springer, New York  (1978).
\bibitem{Kle85} P. Kleinw\H{a}chter and I. Rotter, {\em Phys. Rev.} C {\bf 32}, 1742 (1985).  
\bibitem{Dro00} S. Dro\.zd\.z, J. Oko{\l}owicz, M. P{\l}oszajczak, and I. Rotter, {\em Phys. Rev.} C {\bf 62}, 4313 (2000).
\bibitem{Cha06} R. Chatterjee, J. Oko{\l}owicz, and M. P{\l}oszajczak, {\em Nucl. Phys.} A {\bf 764}, 528 (2006).
\bibitem{Mic07} N. Michel, W. Nazarewicz, and M. P{\l}oszajczak, {\em Phys. Rev.} C {\bf 75}, 031301 (2007).
\bibitem{Oko08} Yan-an Luo, J. Oko{\l}owicz, M. P{\l}oszajczak, and N. Michel, ArXiv:nucl-th/0211068;\\
J. Oko{\l}owicz, M. P{\l}oszajczak, and Yan-an Luo, {\em Acta Phys. Pol.} {\bf  39}, 389 (2008).
\bibitem{Rot00} I. Rotter, E. Persson, K. Pichugin, and P. \v{S}eba, {\em Phys. Rev.} E {\bf 62}, 450 (2000). 
\bibitem{Ber84} M.V. Berry, {\em Proc. R. Soc. London}, Ser. A {\bf 392}, 45 (1984).
\bibitem{Lau94} H.-M. Lauber, P. Weidenhammer, and D. Dubbers, {\em Phys. Rev. Lett.} {\bf 72}, 1004 (1994);
D.E. Manolopoulos, and M.S. Child, {\em Phys. Rev. Lett.} {\bf 82}, 2223 (1999); \\
F. Pistolesi, and N. Manini, {\em Phys. Rev. Lett.} {\bf 85}, 1585 (2000); \\
C. Dembowski {\em et al.}, {\bf 86}, 787 (2001).
\bibitem{Ber84a} M.V. Berry and M. Wilkinson, {\em Proc. R. Soc. London}, Ser. A {\bf 392}, 15 (1984).
\bibitem{Kat95} T. Kato, {\em Perturbation Theory for Linear Operators} (Springer Verlag, Berlin, 1995).
\bibitem{Zir83} M.R. Zirnbauer, J.J.M. Verbaarschot and H.A. Weidenm\"{u}ller, {\em Nucl. Phys.} A {\bf 411}, 161 (1983).
\bibitem{Hei91} W.D. Heiss, and W.-H. Steeb, {\em J. Math. Phys.} {\bf 32}, 3003 (1991).
\bibitem{Duk04} J. Dukelsky, S. Pittel, and G. Sierra, Rev. Mod. Phys. {\bf 76}, 643 (2004).
\bibitem{Duk09} J. Dukelsky, J. Oko{\l}owicz, and M. P{\l}oszajczak, to be published.
\bibitem{DP} R.S. Nikam, P. Ring, and L.F. Canto, Z. Phys. A {\bf 324}, 241 (1986); \\
R.S. Nikam, P. Ring, and L.F. Canto, Phys. Lett. B {\bf 185}, 269 (1987);\\
C. Price, H. Esbensen, and S. Landowne, Phys. Lett. B {\bf 197}, 15 (1987);\\
J. de Boer, C.H. Dasso, and G. Pollarolo, Z. Phys. A {\bf 335}, 199 (1990);\\
Y. Sun, P. Ring, and R.S. Nikam, Z. Phys. A {\bf 339}, 51 (1991);\\
L.F. Canto {\em et al.}, Phys. Rev. C {\bf 47}, 2836 (1993);\\
K.G. Helmer {\em et al.}, Phys. Rev. C {\bf 48}, 1879 (1993).



\end{thebibliography}
\end{document}